# Neural Network Based in Silico Simulation of Combustion Reactions

Jinzhe Zeng[1], Liqun Cao[1], Mingyuan Xu[1], Tong Zhu[1,2,*] and John ZH Zhang[1,2,3,*]

[1] Shanghai Engineering Research Center of Molecular Therapeutics & New Drug Development, School of Chemistry and Molecular Engineering, East China Normal University, Shanghai, 200062, China

[2] NYU-ECNU Center for Computational Chemistry at NYU Shanghai, Shanghai, 200062, China

[3] Department of Chemistry, New York University, New York 10003, United States

Email: tzhu@lps.ecnu.edu.cn, john.zhang@nyu.edu

**Abstract:** Understanding and prediction of the chemical reactions are fundamental demanding in the study of many complex chemical systems. Reactive molecular dynamics (MD) simulation has been widely used for this purpose as it can offer atomic details and can help us better interpret chemical reaction mechanisms. In this study, two reference datasets were constructed and corresponding neural network (NN) potentials were trained based on them. For given large-scale reaction systems, the NN potentials can predict the potential energy and atomic forces of DFT precision, while it is orders of magnitude faster than the conventional DFT calculation. With these two models, reactive MD simulations were performed to explore the combustion mechanisms of hydrogen and methane. Benefit from the high efficiency of the NN model, nanosecond MD trajectories for large-scale systems containing hundreds of atoms were produced and detailed combustion mechanism was obtained. Through further development, the algorithms in this study can be used to explore and discovery reaction mechanisms of many complex reaction systems, such as combustion, synthesis, and heterogeneous catalysis without any predefined reaction coordinates and elementary reaction steps.

Understanding and prediction of chemical reactions are the fundamental demanding in the study of many complex chemical systems such as combustion, catalysis, and synthesis. Limited by the resolution of experimental detection methods, direct experimental characterization of a chemical reaction is extremely challenging. In the past decades, reactive molecular dynamics (MD) simulations have shown their value in providing molecular-level insights into the mechanism of chemical reactions for a series of systems. Compared to experiments, MD simulations can observe all the details of the reaction from different perspectives. In addition, the reaction conditions can be easily controlled in the simulation, some supercritical conditions which are difficult to achieve in the experiment can also be handled.

The heart of the reactive MD simulation is the potential energy surface (PES) which describes the inter- and intramolecular interactions in the system. Currently, there are mainly two classes of methods that can be used to construct the PES of a given molecular system: the quantum mechanics (QM)-based methods and the empirical force fields. Quantum mechanics is undoubtedly more rigorous and accurate, and MD simulations based on it are also known as *ab initio* MD simulation (AIMD)[1]. Unfortunately, the computational cost inherent in the QM calculation severely limits the simulation scale of the AIMD method. Usually, it can only handle simulations of systems containing no more than atoms in a picosecond time scale. With the rapid development of computer hardware and algorithms, especially the employment of graphic processing units (GPUs), some AIMD methods have recently begun to handle larger chemical systems[2]. But so far, it is still unrealistic to use AIMD to simulate large-scale reaction molecular systems.

Compared with QM calculation, the efficiency of empirical reactive force fields such as ReaxFF[3, 4, 5, 6, 7], REBO[8], and MEAM[9] are much higher, making it possible to reach simulation scales that are orders of magnitude beyond what is tractable for conventional AIMD methods[10, 11]. However, efficiency is not a free lunch. Many previous studies have questioned the accuracy of empirical reactive force fields[12, 13, 14]. The most difficult part of developing a reaction force field is the choice of function form, which should accurately describe the various important interactions in the chemical system, especially the breaking and formation of chemical bonds. In addition, the complicated parameterization process also hinders the development of reactive force fields.

Recently, more and more researchers are switching to seek the help of machine learning (ML) methods. ML methods, especially artificial neural networks (NN), provided the possibility to construct PESs with the accuracy of the QM method while its efficiency is as high as to force fields. Neural networks constitute a very flexible and unbiased class of mathematical functions, which in principle is able to approximate any real-valued function to arbitrary accuracy. Since Behler and Parrinello proposed the high-dimensional neural network (HDNN) approach[15, 16, 17, 18, 19, 20], many different NN PESs have been proposed for water, small organic molecules and metal materials. For example, the sGDML and DTNN of Müller et al.[21, 22, 23], the kCON model of Hammer et al.[24], and the DeepMD method of Wang and co-workers[25, 26]. NN potentials have also been employed to study the reaction mechanisms of chemical systems. By combining high-precision NN PESs and quantum dynamics method, Zhang and Jiang's group have studied a series of elementary reactions in the gas phase and on the surface[27, 28, 29, 30]. Liu and co-workers developed the LASP program to study the heterogeneous catalysis with NN PESs[31]. Meuwly et. al also studied the nucleophilic substitution reaction [Cl–$CH_3$–Br]¯ in water with NN potential[32].

To date, NN PESs have only been used to study the reaction mechanisms of small or medium-sized molecular systems, while its performance in large complex reaction systems is not known. In this work, a workflow of preparing reference datasets with a machine learning approach was proposed. Then two NN PESs were trained and used to simulate the oxidization of hydrogen and methane. The potential energy and atomic forces can be accurately predicted by the NN PESs, and benefit from their high efficiency, sub-nanosecond MD trajectories for systems containing hundreds to thousands of atoms were produced. Based on the simulation, detailed combustion mechanisms were obtained and are highly consistent with the experiments.

## Results

**Accuracy of NN PESs.** The performance of the NN potential highly depends on the quality of the reference datasets. Although there are several databases, such as QM7[33], QM9[34], ANI-1[35], and ANI-1x[36], are accessible, they mainly include organic molecules and are therefore not suitable for this work. Combustion will produce a lot of molecular fragments and most of them are free radicals[37]. Therefore, we proposed a workflow (details are listed in the Methodology section) to construct the reference datasets for the combustion. Then the DeepPot-SE model[38] was used to train two NN PESs based on the reference datasets for the hydrogen- and methane-combustion system (For the sake of convenience, we call them $H_2$-system and the $CH_4$-system, respectively). The mean absolute errors (MAEs) and root mean squared errors (RMSEs) of total energies are shown in Table 1. It is clear that DFT energies are accurately reproduced. For the $H_2$-system, the MAE and RMSE are only 0.058eV/atom and 0.099eV/atom in the test set. While for the $CH_4$-system, the errors of energy are slightly larger (the RMSE are 0.14eV/atom and 0.24eV/atom in the training set and test set, respectively.) than that of $H_2$-system, but still very small.

**Table 1.** Energy prediction errors for the hydrogen and methane dataset. The mean absolute errors (MAEs) and root mean squared errors (RMSEs) are in eV/atom.

| System | Training Set / Test Set | | |
|---|---|---|---|
| | Structure Number | MAE | RMSE |
| Hydrogen | 136,910/4,471 | 0.016/0.058 | 0.027/0.099 |
| Methane | 578,731/13,315 | 0.041/0.140 | 0.073/0.240 |

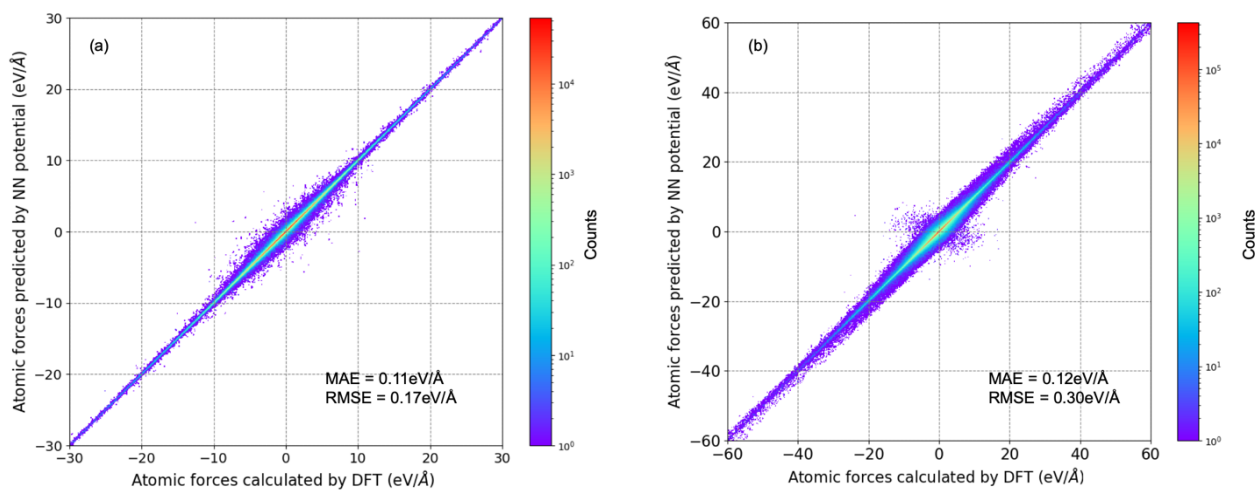

**Figure 1.** The correlation of atomic forces between NN predictions and DFT calculations. (a), results of the $H_2$-system; (b), results of the $CH_4$-system.

As for the atomic forces, the predicted values of the NN models are also highly consistent with the calculated results of the DFT (Fig. 1). The correlation coefficients are 0.999 and 0.999, respectively. And the RMSEs of both systems are within 10 kcal/(mol·Å). Considering that there are a large number of atomic collisions during the combustion, and some atomic forces can be as high as hundreds of eV/Å, the accuracy of these two models is encouraging.

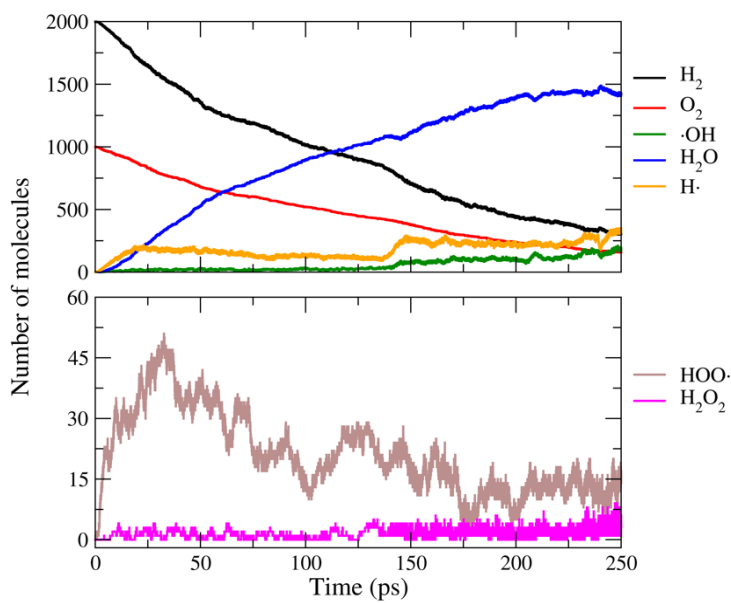

**Figure 2.** Major species generated during the simulation of $H_2$-system.

**The combustion of hydrogen gas.** A 250ps reactive MD simulation was performed for hydrogen combustion with the NN PES by LAMMPS package[39]. The system is a period box contains 2000 $H_2$ molecules and 1000 $O_2$ molecules, whose density is 0.25 g/cm$^3$. NVT MD simulations were performed with a time step of 0.1fs, the temperature was kept at 3000K by using the Berendsen thermostat. Fig. 2 shows the major species generated during the simulation, which include the reactants, two products ($H_2O$ and HOOH), and three active species ·$HO_2$, HO·, and H·. As can be seen, the product is rapidly produced while the reactants are being consumed. After 250ps, most of $H_2$ and $O_2$ are consumed and 1418 water molecules are generated.

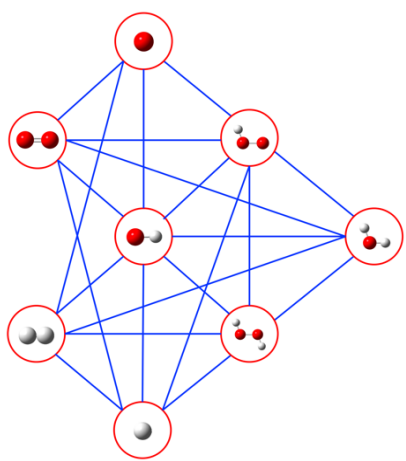

**Figure 3.** Reaction network for hydrogen combustion.

The most important reason for reactive MD simulations of chemical reaction systems is to determine the reaction mechanism. To this end, we analyzed the trajectory with the help of ReacNetGenerator[40]. Fig. 3 shows a reaction network diagram detected from the trajectory. The species types and the relationship between them are basically in line with the experiments[41, 42, 43, 44] and previous theoretical studies[45]. The further careful analysis found that the collision of $H_2$ and $O_2$ can generate H·, O·, HO·, and ·$HO_2$:

$$H_2 + O_2 = 2HO· \tag{r1}$$

$$H_2 + O_2 = H· + ·HO_2 \tag{r2}$$

$$H· + O_2 = HO· + O· \tag{r3}$$

$$O· + H_2 = HO· + H· \tag{r4}$$

Then the HO· and ·$HO_2$ can participate in free radical reactions:

$$2HO· = H_2O_2 \tag{r5}$$

$$·HO_2 + H_2 = H_2O_2 + H· \tag{r6}$$

And water molecules are mainly produced by the following reactions:

$$HO· + H_2 = H_2O + H· \tag{r7}$$

$$HO· + H· = H_2O \tag{r8}$$

$$H_2O_2 + H· = H_2O + HO· \tag{r9}$$

$$·HO_2 + HO· = H_2O + O_2 \tag{r10}$$

These reactions can all be found in the previous theory study[44] and experimental measurments[41, 42, 43, 44], which indicates the success of the NN PESs for the $H_2$-system.

**The combustion of methane.** A 500ps reactive MD simulation was performed for methane combustion with the NN PES. The system contains 100 $CH_4$ molecules and 200 $O_2$ molecules (a total of 900 atoms). The density is also 0.25 g/cm³. Other simulation conditions are the same as that of $H_2$.

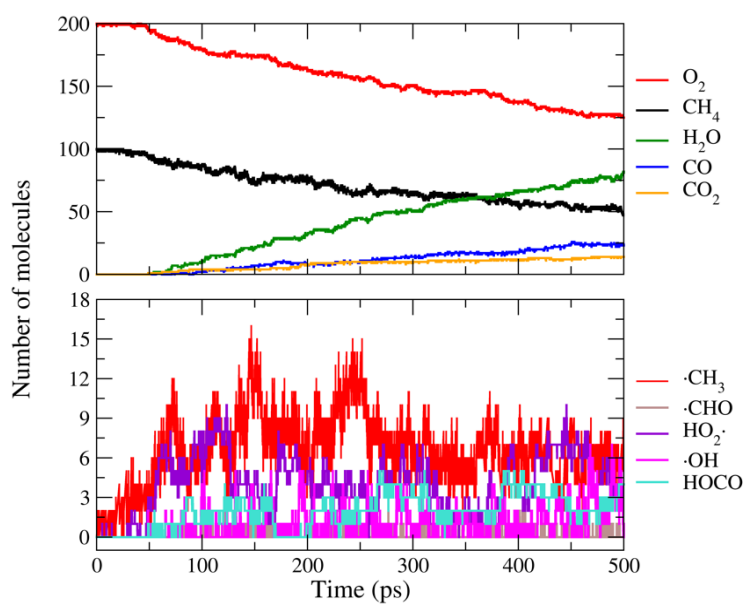

**Figure 4.** Major species generated during the simulation of $CH_4$-system.

Fig. 4 show the changes of the main species during the MD simulation. Similarly, as the $H_2$-system, some expecting (which are important) molecules and free radicals appear in the trajectory. After 500ps simulation, about 50 $CH_4$ and 100 $O_2$ are consumed. More than 140 species and 400 reactions are observed and about 80 $H_2O$, 25 CO, and 15 $CO_2$ are produced. The main reaction routes are shown in Fig. 5. The combustion of methane starts with the abstraction of the H atom from the $CH_4$ by $O_2$, and two free radicals, ·$CH_3$ and ·$HO_2$, will be produced. During the simulation, other radicals such as ·H, ·OH, and ·$HO_2$ can also abstract H atom from $CH_4$ and form ·$CH_3$. The atomization of methane into ·H and ·$CH_3$ was also observed.

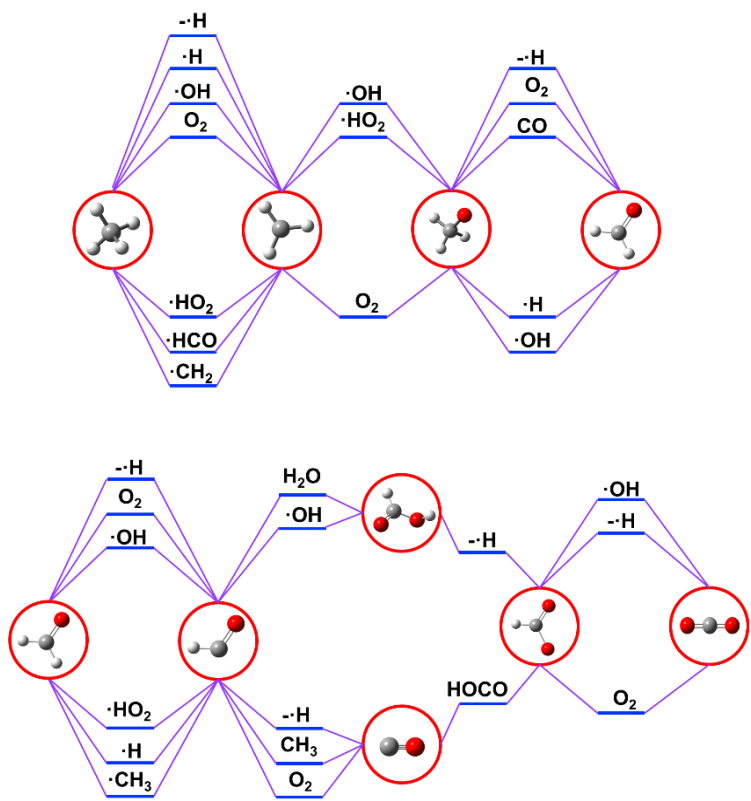

**Figure 5.** Main reaction routes in the simulation methane oxidation with NN PES.

Then the ·CH$_3$ radicals interacted with ·HO$_2$, ·OH and O$_2$ to form the CH$_3$O· radicals. Then the CH$_3$O· radicals were abstracted hydrogen atoms by ·H, ·OH, O$_2$, and CO to form formaldehyde molecules. The CH$_3$O· radicals can also be transformed into ·CH$_2$OH, and then produce formaldehyde by losing a hydrogen atom.

The formaldehydes were further converted into the formyl (HCO·) radicals. The main reaction paths are hydrogen abstraction by ·H, ·OH, O$_2$, ·HO$_2$, and ·CH$_3$. The formyl radical can subsequently lose a hydrogen atom via reactions, thereby forming the carbon monoxide (CO). It can also interact with ·OH and H$_2$O to form the HCOOH molecule and further convert into the ·CO$_2$H radicals, which can be further converted to carbon dioxide. These reactions are consistent with the previous study[37] and experiments[46].

**Discussion**

In this work, two neural networks PESs are trained to describe the complex interactions in the combustion of hydrogen and methane, including the forming and breaking of chemical bonds. These NN models can predict the potential energy and atomic forces of DFT precision, while it is orders of magnitude faster than the conventional DFT calculation. With these two models, reactive MD simulations were performed to explore the combustion mechanisms of hydrogen and methane. Benefit from their high efficiency, nanosecond-sale MD simulations for chemical systems that contain nearly one thousand atoms can be achieved in a week or so. Detailed reaction mechanisms were extracted from the trajectory, which showed excellent agreement with the experiments. Compared to classical force fields, these neural network potentials are not limited by the form of potential function and the complicated parameterization process.

One might concern that a cutoff was used in the training of NN PESs, which means that the interaction between two atoms will be ignored if their distance is larger than the cutoff. This may be a critical issue in condensed matter systems, but gas-phase reactions like combustion are dominated by collisions between molecules, the contribution of long-range interactions to reactions in such a system should be negligibly small (Figure S1.) Another issue to be pointed out is that, although some algorithms were used in this study to minimize the size of the reference dataset, there are still 578,731 structures in the training set of CH4-system, which will consume a lot of computational resources for QM calculation. In order to further minimize the size of the reference set while ensuring its completeness, new algorithms still need to be developed. Many pioneers have come up with some solutions, such as active learning and more complex clustering algorithms, and we are also working on this problem.

However, these issues cannot conceal the bright future of the application of NN-based reactive MD simulation in studying the reaction mechanism of large-scale chemical systems. Through further development, the algorithms in this study can be used to explore and discovery reaction mechanisms of many complex reaction systems, such as combustion, synthesis, and heterogeneous catalysis without any predefined reaction coordinates and elementary reaction steps.

## Methodology

**Reference dataset.** In this study, a workflow was proposed for making reference datasets:

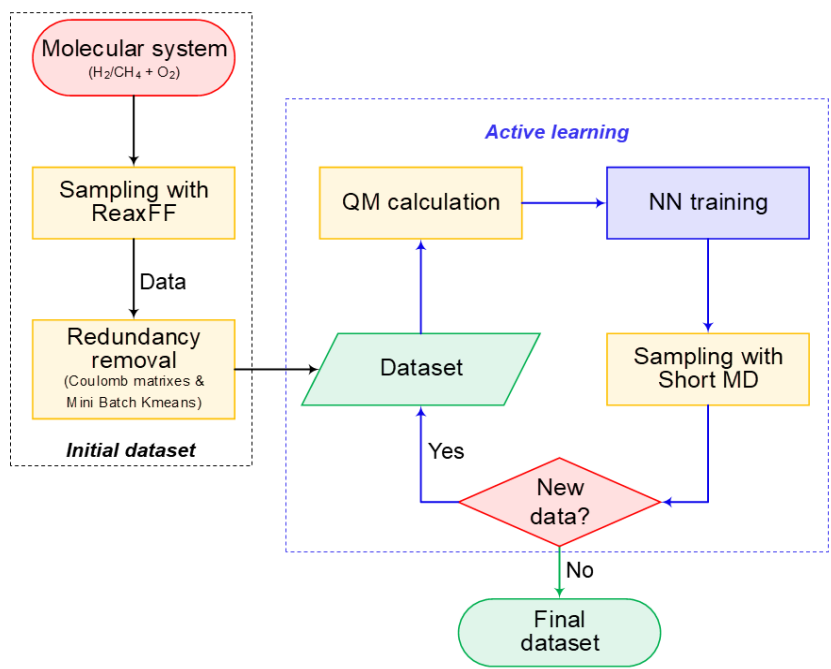

**Figure 6.** The workflow of reference dataset construction.

The details of each module in the workflow are given below.

**Reactive MD simulation with ReaxFF.** Reactive MD simulation with ReaxFF was used to sample the initial dataset. Two molecular systems for the oxidation of hydrogen and methane were built by using the Amorphous Cell module in Material Studio[47]. The hydrogen oxidation system contains 500 hydrogen molecules and 250 oxygen molecules while the methane oxidation system contains 500 methane molecules and 1000 oxygen molecules. The densities of these two systems were set to 0.25g/cm$^3$. The Forcite module in Materials Studio was used to equilibrate these two systems at 298K for 250ps. Then the LAMMPS[48] program was used to perform the MD simulation for 250ps ($H_2$-system) and 2500ps ($CH_4$-system). The NVT ensemble was used and the temperature was set to 3000K with the Berendsen thermostat. The ReaxFF parameter of Chenoweth et al. (CHO-2008 parameter set)[49] was employed. The time step of the simulation was 0.1fs and the atomic coordination was recorded every 10fs.

**The initial dataset.** For the $H_2$- and $CH_4$-system, we got 0.25ns and 2.5ns trajectories, respectively. The Open Babel software[50] and the Depth-First Search algorithm[51] were used to detect species in every snapshot of the trajectory. Then, for each atom in each snapshot, we build a molecular cluster that contains this atom and species that within a specified cutoff centered on it. In this work, the cutoff was set to 5Å. The initial dataset for H$_2$-system contains about 15 million structures, while there are about 22.5 million structures in the CH$_4$-system.

**Redundancy removal.** The initial training sets are too large to perform QM calculations for every molecular cluster it contains. Therefore, it is necessary to resample it to remove redundant structures while ensuring its completeness. To this end, we first classified the initial training sets into sub-datasets based on the chemical bond information of the central atom. For example, the central H atom can be classified into 2 different types: a single H atom (H0) and an H atom formed a single chemical bond with another atom (H1).

Further treatment still needed for large sub-datasets. For a given large sub-dataset, we first expressed each molecular cluster it contains as a Coulomb matrix[52]:

$$C_{ij} = \begin{cases} \frac{1}{2}Z_i^{2.4}, i = j \\ \frac{Z_i Z_j}{|R_i - R_j|}, i \neq j \end{cases} \quad (1).$$

Where $Z_i$ and $Z_j$ are nuclear charges of atom $i$ and $j$, $R_i$ and $R_j$ are their Cartesian coordinates. The minimum image convention[53] was used to consider the periodic boundary condition. "Invisible atoms" were introduced to fix the dimension of the Coulomb matrix. These invisible atoms do not influence the physics of the molecule of interest and make the total number of atoms in the molecule sum to a constant. To lower the dimension of the dataset and keep as much structural information as possible, the Coulomb matrix was further represented by the eigen-spectrum, which is obtained by solving the eigenvalue problem $\mathbf{Cv} = \lambda \mathbf{v}$ under the constraint $\lambda_i \geq \lambda_{i+1}$.

Then the clustering algorithm Mini Batch KMeans[54], was used to cluster the given sub-datasets into smaller clusters according to the eigen-spectrum. Then we randomly selected 10000 structures from each cluster (If the cluster contains no more than 10000 structures, then all of them were selected).

After the redundancy removal process, the dataset of the H$_2$-system contains 136,910 structures while there are 578,031 structures in the dataset of CH$_4$-system.

**Active learning.** Large amplitude collisions and reactions in the combustion can produce a lot of un-predictable species and intermediates. To ensure the completeness of the reference dataset, an active learning approach[55] was used. Four different NN PES models were trained based on the dataset from the last step. Then several short MD simulations were performed. During the simulation, the atomic forces are evaluated by these four NN PES models simultaneously. For a specific atom, if the predicted forces by these four models are consistent with each other, then the molecular cluster which centered on this atom can be found in the dataset. On the contrary, if the results of these four models are inconsistent with each other and the error between them is in a specific range (0.5eV/Å < error < 1.0eV/Å in this work), the corresponding molecular cluster will be added into the dataset. This process will be continued until the predictions of the four models are always consistent.

**QM calculation**. The potential energy and atomic forces for every structure in the final dataset were calculated by Gaussian 16[56] software at the MN15/6-31g** level. The MN15 functional was employed because of its good performance on multiconfigurational systems[57]. To consider the spin polarization effect, the initial wave function of a given structure is obtained by the combination of the wave functions of species that make up it. While the wave functions of the species are obtained based on their charge and spin.

**Training of the NN PES.** The DeepPot-SE (Deep Potential - Smooth Edition) model[38] was used to train the NN potential by DeePMD-kit program[58]. Details of this method can be found in Ref. (57). For both systems, the model includes two networks: the embedding network and the fitting network. Both networks use ResNet architecture[59]. The embedding network was set to (25, 50, 100) and the size of the embedding matrix was set to 12. The size of the fitting network is set to 240, 240, 240 and a timestep is used in the ResNet. The cutoff radius was set to 6.0Å and the descriptors decay smoothly from 1.0 Å to 6.0 Å. The initial learning rate was set to 0.0005 and it will decay every 20000 steps. The loss is defined by

$$\mathcal{L} = p_e \mathcal{L}_e + p_f \mathcal{L}_f \tag{2}$$

where the prefactor of the energy error $\mathcal{L}_e$ is set to 0.2 eV$^{-1}$ and the prefactor of force error $\mathcal{L}_f$ decays from 1000 Å·eV$^{-1}$ to 1 Å·eV$^{-1}$.

**Corresponding Author**

*E-mail: tzhu@lps.ecnu.edu.cn

*E-mail: jz2@nyu.edu

**Acknowledgments**

The author thanks Mr. Linfeng Zhang and Dr. Han Wang for their discussion and help in using DeepPot-SE, DeePMD-kit. T. Zhu would also like to thank Prof. Donghui Zhang for his valuable suggestions in this project.

This work was supported by the National Natural Science Foundation of China (Grants No. 91641116). J. Zeng was partially supported by the National Innovation and Entrepreneurship Training Program for Undergraduate (201910269080). We also thank the ECNU Multifunctional Platform for Innovation (No. 001) for providing supercomputer time.